\documentclass[preprint,a4paper,12pt,floatfix]{revtex4}

\usepackage{amsmath,graphicx,graphics,epsfig,hyperref}
\usepackage{subfig}

\newcommand{\be}{\begin{equation}}
\newcommand{\ee}{\end{equation}}
\newcommand{\ba}{\begin{eqnarray}}
\newcommand{\ea}{\end{eqnarray}}
\newcommand{\Angstrom}{\stackrel{o}{A}}
\newcommand{\sech}{\sec\! {\rm h}}
\newcommand{\arcsinh}{\arcsin \!{\rm h}}

\begin{document}

\title{Splitting of degenerate states in one-dimensional quantum mechanics}

\author{Avik Dutt \footnote{Electronic address: {\em
quantumavik@gmail.com}}
${}^{(1)}$, Trisha Nath\footnote{Electronic address:
{\em trishatinni@gmail.com
}}${}^{(2)}$, Sayan Kar\footnote{Electronic address: {\em
sayan@iitkgp.ac.in}}${}^{(3)}$, Rajesh Parwani\footnote{Electronic
address:{\em parwani@nus.edu.sg}}${}^{(4)}$}
\address{{\rm $^{(1)}$} Department of Electronics and Electrical Communication
Engineering,\\ Indian Institute of Technology, Kharagpur, 721302, India. }
\address{{\rm $^{(2)}$} Department of Physics \& Meteorology, Indian Institute of Technology, Kharagpur
721 302, WB, India. }
\address{{\rm $^{(3)}$} Department of Physics \& Meteorology and Centre for Theoretical Studies, \\ 
Indian Institute of Technology, Kharagpur,
721 302, WB, India.}
\address{{\rm $^{(4)}$} 
Department of Physics and University Scholars Programme,\\
National University of Singapore, Kent Ridge, Singapore.}

%\author{Avik Dutt}
%\email{quantumavik@gmail.com}
%\affiliation{Department of Electronics and Electrical Communication
%Engineering, Indian Institute of Technology, Kharagpur, 721302, India}

%\author{Trisha Nath}
%\email{trishatinni@gmail.com}
%\affiliation{Department of Physics and Meteorology,
%Indian Institute of Technology, Kharagpur, 721302, India}

%\author{Sayan Kar}
%\email{sayan@iitkgp.ac.in}
%\affiliation{Department of Physics and Meteorology \& Center for Theoretical Studies,
%Indian Institute of Technology, Kharagpur, 721302, India}

%\author{Rajesh Parwani}
%\email{parwani@nus.edu.sg}
%\affiliation{Department of Physics and University Scholars Programme,
%National University of Singapore, Kent Ridge, Singapore}

\begin{abstract}
A classic ``no-go" theorem in one-dimensional quantum mechanics can be evaded when the potentials are unbounded below, thus allowing for novel parity-paired degenerate energy bound states.
We numerically determine the spectrum of one such potential and study the 
parametric variation of the transition wavelength
between a bound state lying inside the valley of the potential and another, 
von Neumann-Wigner-like state, appearing above the potential maximum.
We then 
construct a modified potential which is bounded below except when a parameter 
is tuned to vanish.
We show how the spacing between certain energy levels gradually decrease as we 
tune the parameter to approach the value for which unboundedness arises,   
thus quantitatively linking the closeness of degeneracy to the steepness 
of the potential. Our results are generic to a large class of such potentials. Apart from their conceptual interest, such potentials might be realisable in 
mesoscopic systems thus allowing for the experimental study of the 
novel states. The numerical spectrum in this study  is determined using the 
asymptotic iteration method which we briefly review.

\end{abstract}

\maketitle

\section{Introduction}

Bound states in quantum mechanical potential problems usually
have energies below the potential maximum, as most quantum mechanics textbooks remind us.
However, already in 1929 von Neumann and Wigner \cite{neumann} 
first noted the theoretical existence of 
a class potentials which support normalizable bound states above the potential maximum, embedded 
in the continuum of scattering 
states. Later work by Stillinger and Herrick
\cite{stillinger}  provided a more detailed understanding of such 
unusual bound states, indicating their essential quantum mechanical nature.  
The von Neumann--Wigner above--barrier bound states, were 
eventually experimentally verified in the 1990s, by Capasso \emph{et al.} \cite{capasso}, through 
observations on allowed transitions, between a state inside the valley 
and another above the potential maximum, in semiconductor heterostructures.

Recently, potentials unbounded from below, in one dimension, have attracted 
interest because of certain unusual features in their energy 
eigenvalues and  eigenfunctions. The class of potentials studied were 
 non-singular in any finite domain but asymptotically approached negative 
infinity.  
 It was shown that such potentials not only have 
von Neumann--Wigner states, they also exhibit a new feature: {\em parity-paired degenerate} 
bound states with real energies \cite{koleykar,karparwani}, a property 
prohibited for regular one-dimensional potentials 
\cite{Landau}. The study of such potentials is thus conceptually interesting as they highlight how commonly accepted ``no-go" theorems may be evaded.

It should also be noted that such unbounded-below potentials are not just 
mathematical constructs, but have appeared in the 
context of localization of fields 
on the 3--brane in the study of 
the so-called braneworld models with warped extra dimensions in 
high energy physics\cite{rs}. As mentioned in \cite{karparwani}, and discussed below, such potentials might be realisable, to any given degree of accuracy, in actual mesoscopic systems.  

In this paper, we study in detail the spectrum of two potentials so as to 
anticipate future experimental investigations as suggested in 
Ref.\cite{karparwani}. We first numerically compute the eigenvalues, and 
hence transition wavelengths, for the cosh-sech potential to supplement 
the partial information that is available in earlier analytical work \cite
{karparwani}. 

We then modify the unbounded potential in such a way that the 
unboundedness appears only when we tune a parameter
in the potential to vanish. This allows us to study how the degenerate states 
of the unbounded potential split as the parameter is varied, and as one 
moves to the more realistic bounded potential. Since in \cite{karparwani} we had shown the existence of large classes of potentials supporting degenerate bound states, the splitting of the degenerate states for the bound versions of 
those potentials is naturally expected and thus our results are qualitatively 
generic.  

The eigenvalues are determined with the help of the asymptotic iteration method (AIM), a relatively novel technique to solve second-order linear 
homogeneous differential equations with variable coefficients \cite{cifti,
fernandez,aimappl1,aimappl2,aimappl3} (details of the method are provided in Appendix A). Its asset is that it 
 is able to easily 
circumvent some of the profound analytical and numerical difficulties 
posed for the Schr\"odinger equation with 
potentials which are either singular or unbounded from below.

Our article is organised as follows. In Section II, we discuss the cosh-sech 
potential and determine its spectrum using the asymptotic iteration 
method (AIM).
In Section III we discuss the degeneracy of states and how the levels split 
in the modified bounded potential that we construct.
Section IV summarises our results while the appendices (Appendix A and Appendix B) contain some 
technical details about AIM and physical units, respectively. 

\section{A modified P\"osch--Teller potential}
Consider the cosh-sech potential,
\be
V(x) = -\frac{b^2}{4}\cosh^2x-\left(a^2-\frac{1}{4}\right)\sech^2 x
\label{eq:cosh-sech1}
\ee
This differs from the potential studied in \cite{sous} just 
by an additive constant, $-b^2/4$, since they \cite{sous} 
choose the first term as a $\sinh$ function. 

%The potential is a symmetric inverted potential. For $b<\sqrt{4a^2-1}$, the potential is like an inverted double well, while for $b \ge\sqrt{4a^2-1}$, it is like an inverted oscillator with one maximum on either side of the local minimum at $x=0$.

The above potential has a maximum at 
\[ x_{max} = \cosh^{-1}\left( \frac{4a^2-1}{b^2}\right)^{1/4} \]
which exists only for $b<\sqrt{4a^2-1}$ since $\cosh(x)\geq 1$. The extrema of the potential function are $
%\begin{eqnarray}
V_{max}=-\frac{b}{2}\sqrt{4a^2-1}\label{eq:max}$ and $V_{min}=-\left(a^2-\frac{1}{4}\right)- \frac{b^2}{4}\label{eq:min}.
%\end{eqnarray}
$
The potential is of the same form as the volcano potential of  Ref.~\cite{koleykar},
\be
V(x) = -A_1 \cosh ^{2\nu} x -A_2 \sech ^2 x
\label{eq:cosh-sech2}
\ee
with $\nu=1$, which can be generated from the general ansatz of Ref.~\cite{karparwani} by choosing $f(x) = \cosh^{-\nu/2} x$. 

To solve the time--independent Schrodinger equation using the AIM 
(see Appendix A), 
the wave function is written as $\psi=e^{-f(x)} y$ \cite{sous}, where
\begin{equation}
    f(x)=ib\sinh(x)+(a-1/2)\ln \cosh(x) \; 
\end{equation}
A further change of independent variable to $u=\sinh x$, transforms the 
Schrodinger equation to 
\begin{equation}
    y''=\lambda_0(u)y'+s_0(u)y
\end{equation}
where
\begin{equation}
    \lambda_0(u)=2\left[i b+\frac{a-1}{u^2+1}u\right]
\end{equation}
\begin{equation}
    s_0(u)=-\frac{2iub(a-1)}{u^2+1}-\frac{E}{u^2+1}+\frac{a-a^2-1/4}{u^2+1}+\frac{3b^2}{4}
\label{eq:s0}
\end{equation}
The above form of the Schrodinger equation is necessary in order to use AIM, 
to find energy eigenvalues for
different $a, b$ values.

\subsection{Energy spectra and transition wavelengths}
We now present the results of our calculations using the asymptotic 
iteration method. All computations were carried out using standard routines 
in {\em Mathematica}
with default precision.

In Fig.~\ref{fig:energies}, we plot the dimensionless energy eigenvalues
(see Appendix B for a discussion on physical units) 
for the unbounded potential with $a=20$ and $b$ varying from $0$ up to $10$. 
One can observe that 
for energy eigenvalues, as $b$ increases, energies increase in general. 
Our calculations (not shown here) also show that the energy eigenvalues 
monotonically decrease 
with the value of the parameter $a$. %From $0.5 \, to \,10$ L1 is in the visible range and from $1.5 \, {\rm to} \,5$, others are in the visible range.

\begin{figure}
\centering
%\subfloat[Energy eigenvalues for the cosh-sech potential, $a=20$.]
\includegraphics[width=.98\textwidth]{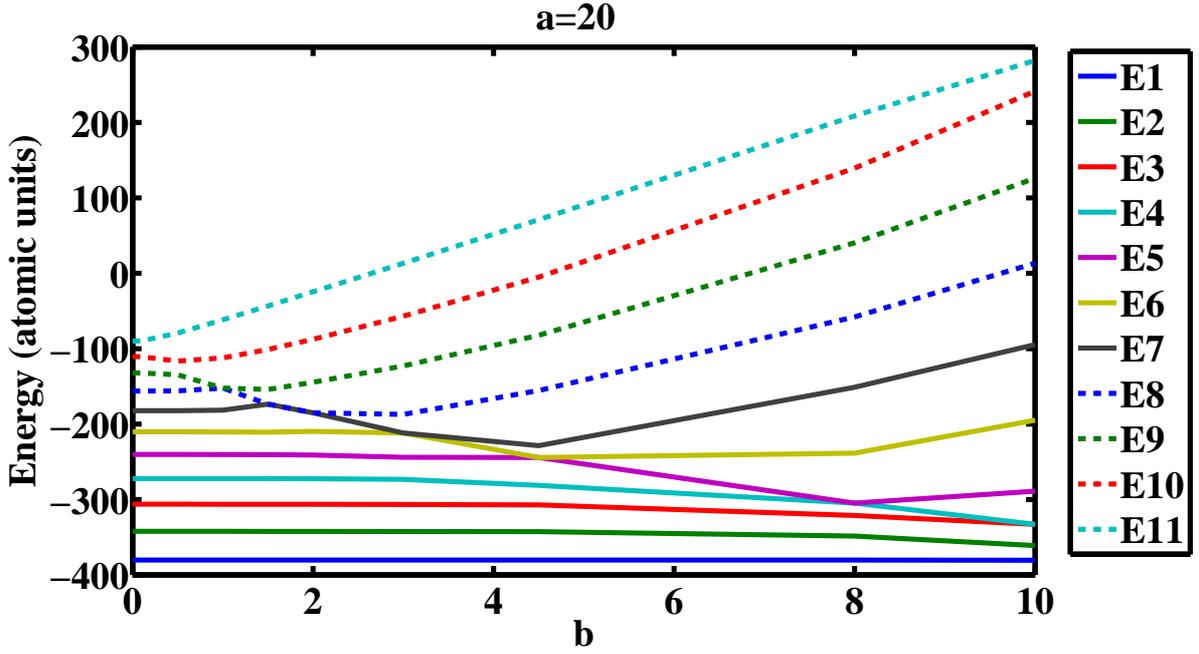}
%\subfloat[Wavelength of transition between adjacent energy levels.] %{\label{fig:wavelength}\includegraphics[width=.48\textwidth]{graphics/wavelengthscoshsech}}
\caption{Variation of energy eigenvalues with $b$ for the cosh-sech potential for $a=20$.}
\label{fig:energies}
\end{figure}

% Fig.~\ref{fig:a} shows the variation of the dimensionless energy eigenvalues and the transition wavelengths for the cosh-sech potential with $a$ for $b=0.10$. Each wavelength and energy value decreases as $a$ increases. All wavelengths are in the visible range for $a$ varying from $30 \, {\rm to} \,50$.
%
%\begin{figure}
%\centering
%%\subfloat[Energy eigenvalues for the cosh-sech potential, $a=20$.]
%\includegraphics[width=.98\textwidth]{graphics/energyvsa}
%%\subfloat[Wavelength of transition between adjacent energy levels.] {\label{fig:wavelengthvsa}\includegraphics[width=.48\textwidth]{graphics/lambdavsa}}
%\caption{Variation of energy eigenvalues for the cosh-sech potential with the parameter $a$ for $b=0.10$.}
%\label{fig:a}
%\end{figure}

\begin{figure}
\centering
\includegraphics[width=.98\textwidth]{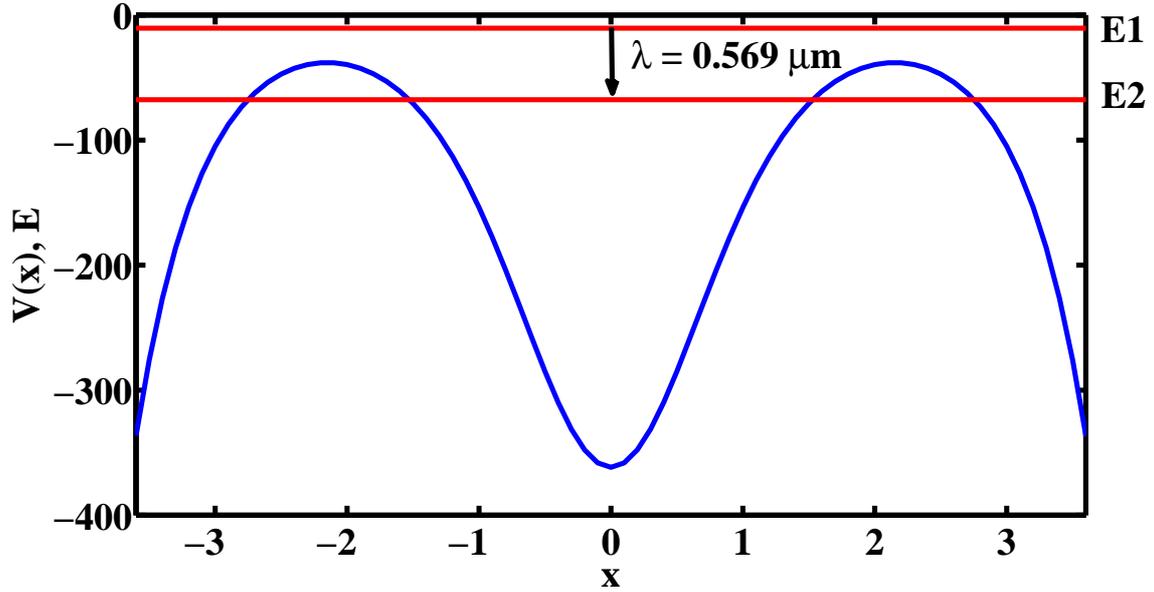}
\caption{Schematic of the cosh-sech potential and the transition wavelength between a pair of states E1 and E2 (in red) located outside and within the potential well (in blue) respectively, for $a=19, b=2$. The wavelength $\lambda = hc/(\tilde E1-\tilde E2) =0.569\, \mu m$ lies in the visible range of the spectrum. The energy scale, the potential and the x-coordinate are plotted in atomic units.}
\label{fig:trans}
\end{figure}

\begin{figure}[!h]
\centering
\includegraphics[width=.98\textwidth]{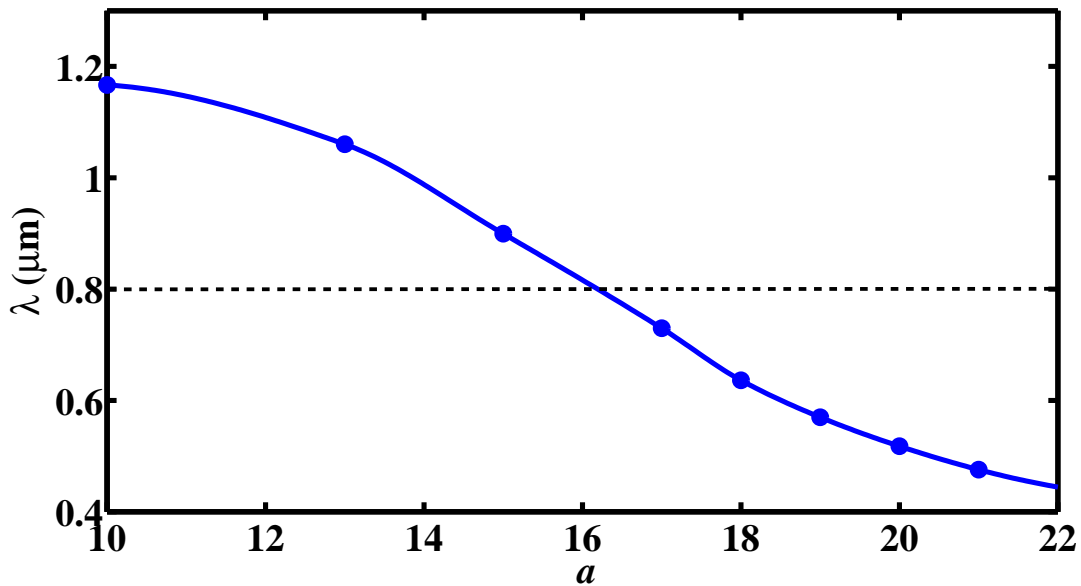}
\caption{The variation of the transition wavelength between a pair of states outside and inside the cosh-sech potential well with the parameter $a$, for $b=2$. The black dashed line represents the rough boundary between the visible and the infra-red regions of the optical spectrum.}
\label{fig:transvsa}
\end{figure}

\begin{figure}[!h]
\centering
\includegraphics[width=.98\textwidth]{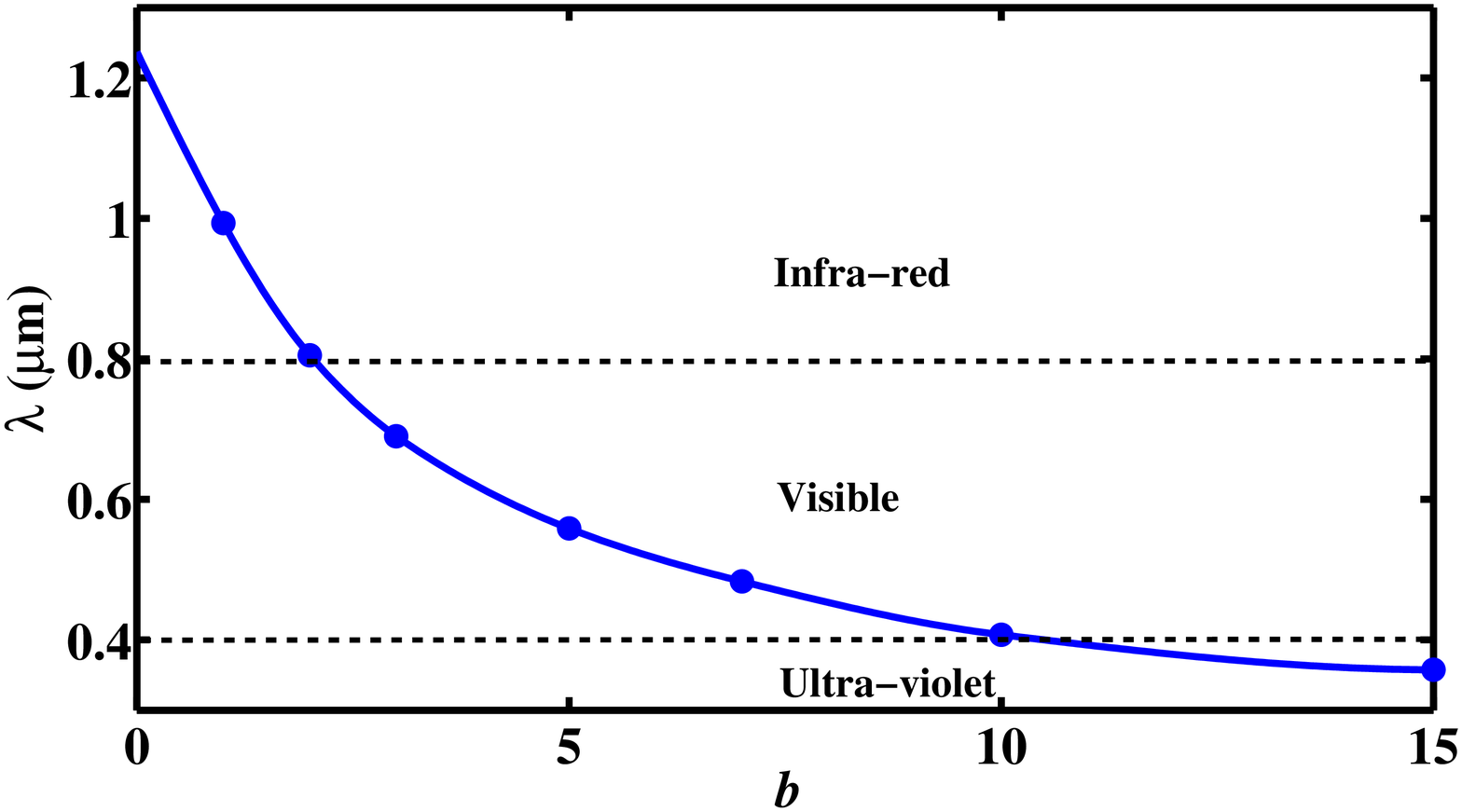}
\caption{The variation of the transition wavelength between a pair of states outside and inside the potential well with the parameter $b$, for $a=16$. The black dashed lines represent the rough boundary between the infra-red, the visible and the ultraviolet regions of the optical spectrum.}
\label{fig:transvsb}
\end{figure}

The transition wavelength between a pair of states can be determined from the difference in their  energies, after converting to suitable physical units, as outlined in Appendix B. Fig.~\ref{fig:trans} demonstrates a pair of eigenstates 
located above (E1) and below (E2) the maximum in the potential well. The 
parameters $a$ and $b$ of the cosh-sech potential have been chosen so that 
the transition wavelength lies in the visible range of the electromagnetic spectrum. 

The transition wavelength can be varied over the range of interest by modifying the parameters $a$ and $b$;  from Fig.~\ref{fig:transvsa}, it is evident that for $a=10$ to 16, $\lambda$ is in the IR region, and greater values of $a$ lead to transitions in the visible region. Similarly, Fig.~\ref{fig:transvsb} shows a variation of the transition wavelength from the IR to the ultra-violet region by changing the parameter $b$. Note that since the energy differences increase with an increase in $b$, hence $\lambda$ moves from the longer IR wavelengths to the relatively short visible region.

For illustration, the length scale parameter $x_0$ (see Appendix B)
had been chosen to be 10 $\Angstrom (= 0.001\, \mu m)$ in the above 
calculations. $x_0$ can be modified suitably to obtain orders of magnitude 
variation in the energy eigenvalues (in physical units) and the 
corresponding transition wavelengths.
  
\section{Degeneracy}
As mentioned earlier, a  curious feature of the class of unbounded potentials being considered here is the existence of degeneracy of the bound eigenstates despite the fact that we are dealing with one-dimensional quantum mechanical problems \cite{koleykar, karparwani}. Not all potentials unbounded from below as $x\to \pm \infty$ support degenerate bound states; for instance the quartic anharmonic oscillator with $V(x) = x^2-Ax^4$, $A>0$ is not known to exhibit degeneracy.

Here we show, numerically, that the degeneracy exists in our potential for many excited states which lie above the previously analytically obtained pair of states (possibly ground states) in Ref.\cite{karparwani}. 

Consider the cosh-sech potential with $a=3, b=1$. After 12 iterations, 
the following eigenvalues are obtained using the AIM: $-6.47301,-6.3402, -2.6222,-2.6058,0.9607,0.9624$. It can be inferred that the energy eigenvalues occur 
in closely separated pairs. Moreover, the separation between the two 
eigenvalues in a pair keeps on decreasing with an increase in the number of 
iterations, suggesting that they are ideally degenerate. The energy eigenvalues below the minimum in $V(x)$ at $x=0$ do not correspond to bound states. 
This is discussed in \cite{choho} (see also \cite{sous}), where the authors
show how the below-minimum states are not linked to  
total--transmission (TT) modes, unlike genuine bound states which are
always constructed out of the TT modes. Further, we have also seen that
these below-minimum states do not occur in pairs, a feature different from
the above-minimum bound states. Thus, in all our
evaluations using AIM we have ignored such states.
%
%\begin{figure}
%\centering
%\includegraphics[width=.98\textwidth]{a3b1VEx}
%\caption{Potential and energy for $a=3, b=1$ after 12 iterations of the AIM. The energy eigenvalue below the minimum in the potential is non-degenerate, whereas the ones above the minimum are degenerate.}
%\label{fig:degen1}
%\end{figure}

\begin{figure}
\includegraphics[width=.98\textwidth]{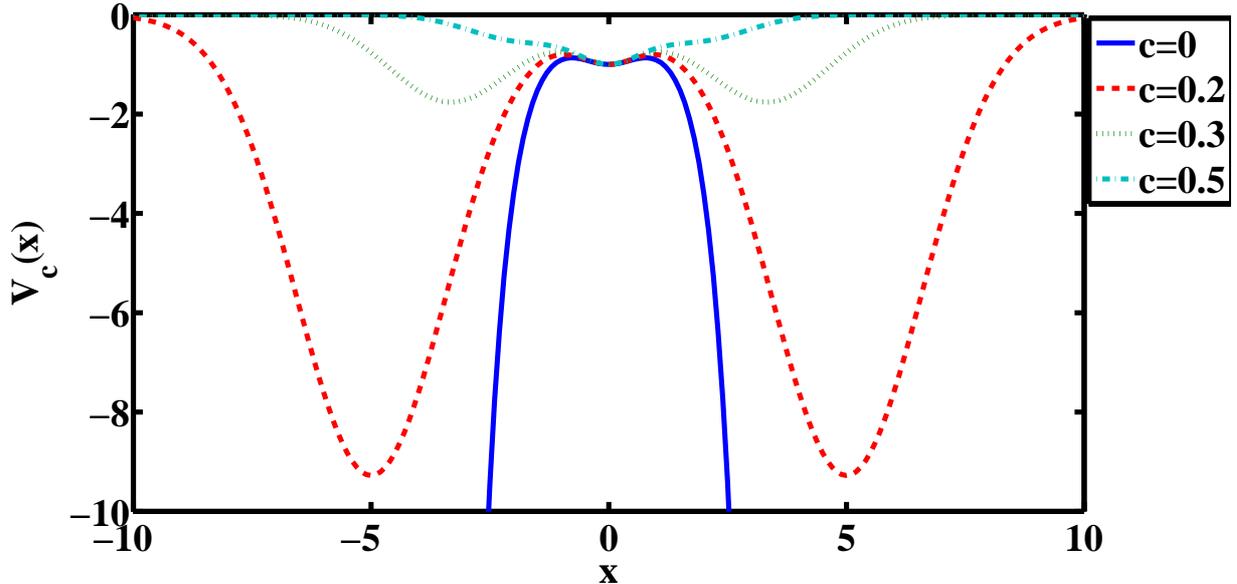}
\caption{The gradual modification of the cosh-sech potential $V_c(x)$ (Eq.~\ref{eq:vcx}) with the parameter $c$ for $a=b=1$. The parameter $c$ can be used to tune the unboundedness of the potential.}
\label{fig:vcx}
\end{figure}

Next, we consider a modified potential which diminishes to zero instead of going to infinity as $x\to\pm\infty$, and which should be realisable in a laboratory. We multiply the cosh term in Eq.\ref{eq:cosh-sech1}, which is responsible for the unboundedness, by an exponentially decaying term controlled by the parameter $c$. 
\be
V_c(x)=-\frac{b^2}{4}\cosh^2x\, \times\exp(-cx^2)-\left(a^2-\frac{1}{4}\right)\sech^2x
\label{eq:vcx}
\ee
The original potential can be recovered by setting $c=0$. As $c$ increases, the first term gets suppressed, and the potential retreats to zero at a faster rate. This effect is illustrated in the Fig.~\ref{fig:vcx} for $a=b=1$. For $c=0.20$, the behaviour of $V_c(x)$ is quite close to that for $c=0$ near the origin, but far from $x=0$, the potential decays to zero instead of the latter case which goes to $-\infty$. This offers promise in investigating the relationship between unboundedness and degeneracy. On the other hand, the $c=0.5$ case is similar to the bounded potential well or the P\"osch-Teller potential rather than the unbounded cosh-sech potential. Thus, $c$ can be used to tune the steepness of the potential away from the origin. 

The energy spectrum of the modified potential $V_c(x)$ can be obtained via the AIM by changing the expression for $s_0(u)$ in Eq.~\ref{eq:s0} to:
\be
s_0(u) = -\frac{2iub(a-1)}{u^2+1}-\frac{E}{u^2+1}+\frac{a-a^2-1/4}{u^2+1} 
+ \frac{b^2}{4}(2+\exp(-c\ \arcsinh^2u))
\ee
The $\arcsinh(u)$ factor occurs in the exponent due to the intermediate variable transformation $u=\sinh x$.

\begin{figure}[!h]
\includegraphics[width=.98\textwidth]{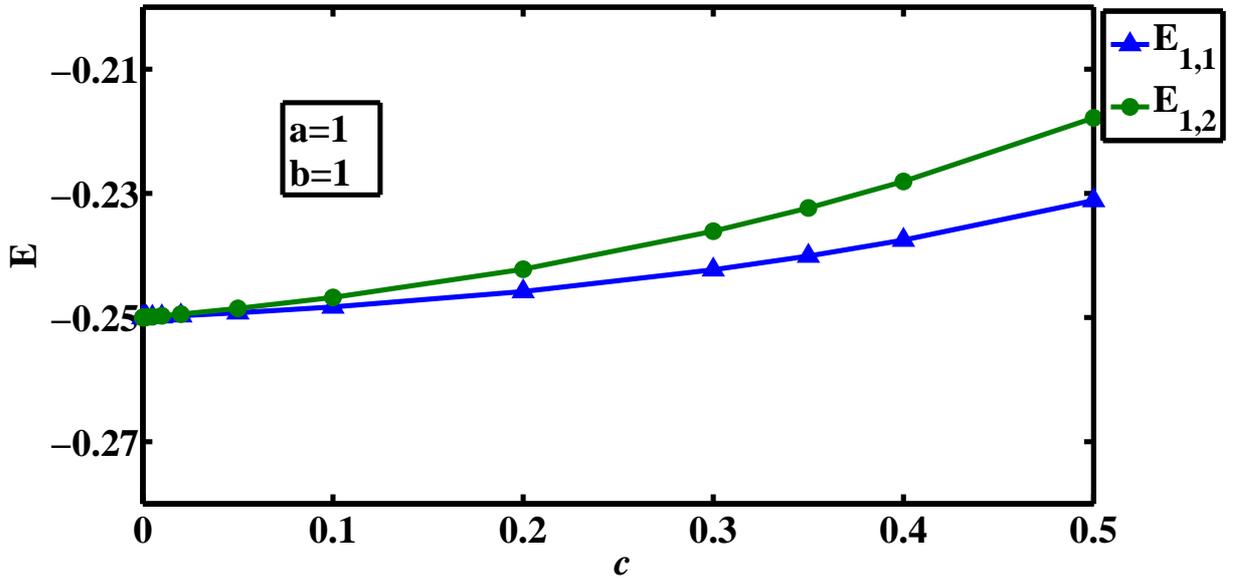}
\caption{Splitting of degeneracy with an increase in the parameter $c$. As the potential varies, the difference in energy between the two eigenstates 
increases.}
\label{fig:vcxenergy}
\end{figure}

\begin{figure}[!h]
\includegraphics[width=.98\textwidth]{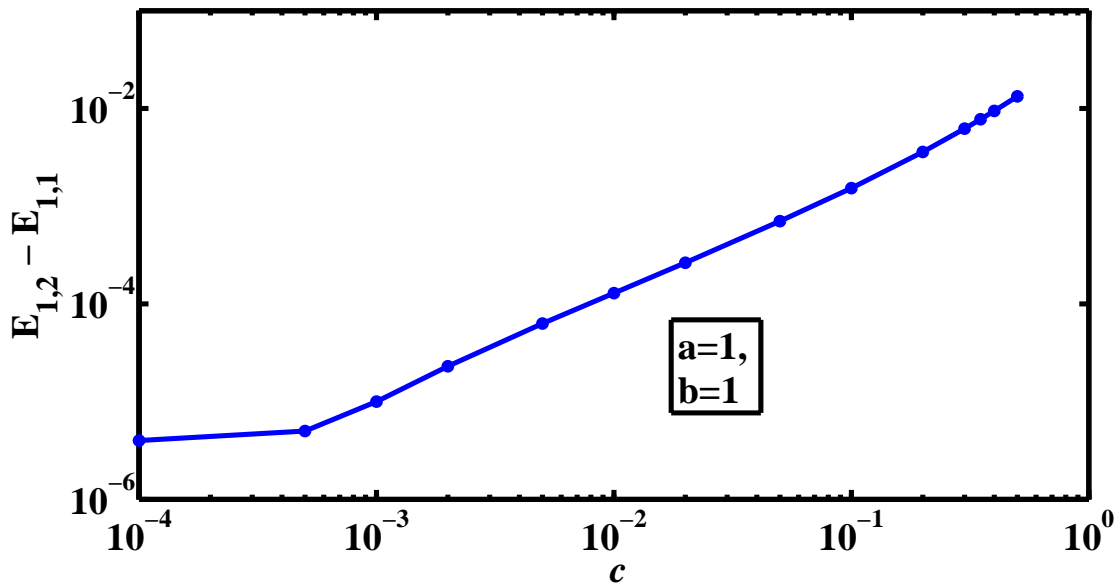}
\caption{Variation of the difference $\Delta E$ as a function of $c$. The
axes are logarithmic.}
\label{fig:diffElog}
\end{figure}

Fig.~\ref{fig:vcxenergy} shows the energy eigenvalues of the pair of degenerate states with the parameter $c$ for the specific case of the potential considered above (viz. $a=b=1$). As we have seen before, a bound state exists for the unbounded potential at -0.25. It is evident from the plot that the pair of degenerate levels splits into separate levels with an increase in $c$. The splitting
is also shown separately in Fig.~\ref{fig:diffElog} where the energy
difference is plotted. We note that as we move towards the left 
(i.e. $c\rightarrow 0$), the minute difference between the two degenerate 
eigenvalues which still remain even at $c=0$, is due to numerical error 
in the AIM rather than actual splitting. This is why the curve is steadily 
increasing for $c > 5\times {10}^{-3}$ and flattens for $c < 5\times
{10}^{-3}$. In fact, $E_{1,2} - E_{1,1}$ does not go below $\sim 5\times
{10}^{-6}$ even for $c=0$, if we use 12 iterations. 

Thus, tuning the parameter $c$ modifies the potential from an unbounded to a bounded one, and gradually increases the separation between the levels which are degenerate in the unbounded case. While it might be impossible to physically realize the unbounded potential ($c=0$), its variants such as the potentials $V_c(x)$ with $c>0$ can be experimentally fabricated through methods of band-gap engineering in semiconductor heterostructures \cite{capasso1}, 
where potentials with multiple
wells and barriers do arise and are not very uncommon. The experimental 
energy eigenvalues can thus be compared with the theoretically predicted 
results with the help of measured transition wavelengths. 

\section{Conclusion}

In this paper we first calculated 
the energy spectra and transition wavelengths between states with energies 
lying within the potential valley and above the potential 
maximum for the volcano potential (\ref{eq:cosh-sech1}), thus extending in detail the analytically obtained results in \cite{karparwani}. 

We emphasize that we have been able to demonstrate, through our numerical work in this paper, 
that there exist several degenerate excited states in the full spectrum, beyond 
the single pair obtained analytically in 
Ref.\cite{karparwani}. We then constructed a new potential, depending on a 
tunable parameter $c$, which was similar to the volcano  
potential near the origin  but which was bounded as $x\rightarrow \pm \infty$ for $c \neq 0$. This non-singular potential exhibited non--degenerate energy  eigenvalues. For $c=0$ one has the original volcano potential with degenerate states and as $c$ is increased, creating a bounded potential, the splitting of degenerate states also increased. 

As stated in Section I, the properties of the bounded potential we studied here are likely to be qualitatively similar to others that can be constructed from the general ansatz of Ref.\cite{karparwani} and plausibly amenable to experimental realization in 
semiconductor heterostructures.  We believe that
laboratory fabrication of systems where 
such potentials may arise
could be a route to the observation of the novel parity paired (almost) degenerate 
states in such one-dimensional problems. Further, the splitting of
the degeneracy will lead to tunable (varying $c$) long wavelength transitions between the
closely spaced split levels for the bounded potential. As for the case of 
the von Neumann-Wigner states, we feel that the study of such parity-paired 
degenerate states is of intrinsic conceptual interest.

Finally, the present study once again shows that the AIM is a useful
semi-analytical tool 
to solve Scr\"odinger-type eigenvalue problems. 
We have verified the pre-existing 
bound state analytical solutions for unbounded potentials, with the help of 
the AIM, and extended the method to determine the entire bound state 
energy spectrum of these potentials which are unbounded below.

\section*{Appendix A: The asymptotic iteration method}
The asymptotic iteration method (AIM) \cite{cifti} is a general semi-analytical technique for solving second order, linear homogeneous ordinary differential equations (ODEs). It has been used quite extensively over the last few years,
in a variety of contexts (see \cite{aimappl1,aimappl2,aimappl3} for a few references). 
The basic idea behind the method is as follows. 
Let $\lambda_0(x)$ and $s_0(x)$ be functions defined over a certain interval, 
having appropriately many successive derivatives. The differential equation
we are concerned with is given as
\be
\frac{d^2 y }{dx^2} = \lambda_0 (x) \frac{dy}{dx} + s_0(x) y
\label{eq:ODE}
\ee
This equation is to be solved by the AIM. 
The technique consists of iteratively 
differentiating the equation to obtain an ODE of the same general type, 
where the coefficients of the next step $\lambda_k(x),\ s_k(x)$ are related to the previous coefficients by the expressions
\ba
\lambda_k(x) &=& \frac{d}{dx}\lambda_{k-1}(x)+s_{k-1}(x) + \lambda_0(x) \lambda_{k-1}(x) \label{eq:aim1}\\
s_k(x) &=& \frac{d}{dx}s_{k-1}(x) + s_0(x) \lambda_{k-1}(x). \ \ \ (k \in { I\!\!N}) \label{eq:aim2}
\ea
The iteration is terminated when the ratio between the coefficients becomes independent of the index $k$,
\be
\frac{s_k(x)}{\lambda_k(x)} = \frac{s_{k-1}(x)}{\lambda_{k-1}(x)} =\alpha(x)
\label{eq:alpha}
\ee
in which case the equation is exactly solvable with the help of the AIM and exact eigenvalues are obtained by finding the roots of the discriminant equation
\be
\delta_k(x) = s_{k-1}(x) \lambda_k(x) - s_k(x) \lambda_{k-1}(x) = 0
\label{eq:quant}
\ee
On the other hand, if the ratio of succesive coefficients is not strictly independent of $k$, exact solutions cannot be found by the AIM. Nonetheless, an approximation to the eigenvalues can be obtained by forced imposition of the condition in Eq.~\ref{eq:alpha} for a sufficiently large value of $k$, wherein lies the asymptotic nature of the method. 
%This is a more general case, since the class of non-exactly solvable problems and quasi exactly solvable (QES) problems far exceeds the number of exactly solvable problems.

Furthermore, after some algebra, it is shown in \cite{cifti} that the solution $y(x)$ to Eq.~\ref{eq:ODE} can be obtained by the following integral relation:
\be
y(x)=\exp\left(-\int^x\alpha dt\right)\left[C_{2}+C_{1}\int^x\exp\left(\int^t \left(\lambda_{0}(\tau)+2\alpha(\tau)\right)d\tau\right)dt\right]
\ee
The time independent Schr\"odinger equation is of the form
\be
\frac{d^2\psi}{dx^2}+\frac{2m}{\hbar^2}(E-V(x))\psi = 0
\label{eq:schrodinger}
\ee
By a suitable transformation of $\psi(x)$ which reflects the asymptotic behaviour of the wave function, Eq.~\ref{eq:schrodinger} can be recast into a form resembling Eq.~\ref{eq:ODE}, where $\lambda_0(x)$ and $s_0(x)$ depend on the energy $E$. Applying the asymptotic iteration method, the $\lambda_k$'s and $s_k$'s are determined to arrive at the ``quantization" condition of Eq.~\ref{eq:quant}. Since $\delta_k$ depends on both $x$ and $E$, it becomes necessary to choose a suitable value of $x=x_0$ in order to solve for the energy eigenvalue $E$. For exactly solvable models, an arbitrary choice of $x_0$ suffices to arrive at the correct eigenvalue. On the contrary, for models which are not exactly solvable, the choice of $x_0$ affects the rate of convergence of the method, and hence the accuracy of the energy calculated.

It is noteworthy that the AIM is essentially an analytical technique, as the successive derivatives required are determined by analytic or symbolic differentiation. However, the form of $\delta_k(x)$ is usually so complicated that the use of numerical algorithms is necessary in order to calculate the roots of the 
equation $\delta_k=0$, with a choice of $x=x_0$ where $x_0$ is usually taken
to be the critical (maximum/minimum) point of the potential.

We tested the AIM on the potentials mentioned in \cite{karparwani} and verified that the numerical eigenvalues agree with the partial information already available analytically. For example, 
we have checked that for $a=1$  there is always one 
energy eigenvalue  $E=-0.25\ \forall\, b$, which corresponds to the known exact 
solution \cite{karparwani}. New results are presented in the main body of this paper.

\section*{Appendix B: Physical units}
In dimensionless (atomic) units, the time independent Schr\"{o}dinger equation is:
\be
\left\{-\frac{d^2}{dx^2}+V(x)\right\}\psi(x)=E_n\psi(x)\label{eq:nodim}
\ee
where $x$ is a dimensionless variable proportional to the one-dimensional spatial coordinate and $V(x)$ is the potential proportional to that in the actual quantum mechanical problem.

On the other hand, the Schr\"{o}dinger equation in real units is:
\be
\left\{-\frac{\hbar^2}{2m}\frac{d^2}{dr^2}+\tilde V(r)\right\}\tilde\psi(r)=\tilde E_n\tilde\psi(r)\label{eq:dim}
\ee
The dimensionless coordinate $x$ can be related to the dimensional coordinate $r$ with the help of a length scale parameter $x_0$, $x=r/x_0$. Here, $x_0$ is a measure of the width of the potential. Substituting this relation in Eq. \ref{eq:dim}, we get:
\be
-\frac{d^2\psi}{dr^2} + \frac{2mx_0^2}{\hbar^2}\tilde V(x) \psi(r)=\frac{2mx_0^2}{\hbar^2}\tilde E_n\psi(r)
\ee
Comparing with Eq. \ref{eq:nodim}, 
\be
\frac{2mx_0^2}{\hbar^2}\tilde V(r) = V(x),\ 
\frac{2mx_0^2}{\hbar^2}\tilde E_n = E_n
\ee
Thus, we have,
\be
\tilde E_n {\rm(real\ units)} = \frac{\hbar^2}{2mx_0^2} E_n (\rm no\ dimensions)
\ee

For example, an electron with a mass $m_e=9.11\times 10^{-31} kg$ has an energy:
\begin{eqnarray}
\tilde E_n ({\rm real \ units}) &=& \frac{6.1042 \times 10^{-39} E_n ({\rm no\ dim.})}{x_0^2} J\nonumber\\
\ &=&\frac{3.8104\, eV}{x_0^2\ ({\rm in \Angstrom})}  E_n ({\rm no\ dim.})
\end{eqnarray}
If we take $x_0 = 10 \Angstrom$, the dimensionless energy is related to the real energy by the factor,
\be
\tilde E_n ({\rm real \ units}) = 38.104\  meV \times E_n\ \   (= 6.1042\times 10^{-21} J \times E_n)
\ee
Corresponding to an energy difference $\Delta\tilde E$, the transition wavelength $\lambda\ ({\rm in\ \mu m})=hc/\Delta \tilde E\approx 1.2424/\Delta\tilde E (\rm in\ eV)$. As an example, consider the red and violet ends of the visible 
spectrum corresponding to wavelengths of 0.8 $\mu m$ and 0.4 $\mu m$, which 
translate to dimensionless energy differences of 81.51 and 40.75 respectively. 
By suitable choice of the parameters in the potentials, it is possible to 
obtain transition wavelengths in the optical range or beyond.

\end{document}